\documentclass[12pt]{article}
\usepackage{amssymb,bbold}
\usepackage{epsfig}
\usepackage{bm}

\newcommand{\sect}[1]{\setcounter{equation}{0}\section{#1}}

\newcommand{\e}{\epsilon}
\newcommand{\bea}{\begin{eqnarray}}
\newcommand{\eea}{\end{eqnarray}}
\newcommand{\nn}{\nonumber \\}
\newcommand{\p}[1]{(\ref{#1})}

\def\be{\begin{equation}}
\def\ee{\end{equation}}
\def\ba{\begin{eqnarray}}
\def\ea{\end{eqnarray}}

\font\mybb=msbm10 at 11pt 
\def\bb#1{\hbox{\mybb#1}}

\def\bR {\bb{R}}

\def\cL{{\cal L}}

\def\cF{{\cal{F}}}

\def\e{\epsilon}

\def\c{\chi}

\begin{document}

\baselineskip 18pt

\begin{titlepage}

\vfill
\begin{flushright}
\today\\
QMUL-PH-03-04\\
hep-th/0304064\\
\end{flushright}

\vfill

\begin{center}
{\bf \Large All supersymmetric solutions of minimal gauged }\\
\vspace*{3mm} {\bf \Large supergravity in five dimensions}

\vskip 10.mm {Jerome P. Gauntlett$^{1}$ and Jan B.
Gutowski$^{2}$}
\\

\vskip 1cm

{\it
Department of Physics\\
Queen Mary, University of London\\
Mile End Rd, London E1 4NS, UK
}\\

\vspace{6pt}

\end{center}
\par

\begin{abstract}
\noindent All purely bosonic supersymmetric solutions of
minimal gauged supergravity in five dimensions are classified.
The solutions fall into two classes depending on whether
the Killing vector constructed from the Killing spinor is
time-like or null. When it is timelike, the solutions
are determined by a four-dimensional K\"ahler base-manifold,
up to an anti-holomorphic function, and generically preserve
1/4 of the supersymmetry. When it is null we provide a precise
prescription for constructing the solutions
and we show that they generically preserve 1/4
of the supersymmetry.  We show that $AdS_5$ is the unique
maximally supersymmetric configuration.
The formalism is used to construct some new solutions,
including a non-singular deformation of $AdS_5$,
which can be uplifted to obtain new solutions of type IIB supergravity.

\end{abstract}
\vskip 1cm
\vfill \vskip 5mm \hrule width 5.cm \vskip 5mm {\small
\noindent $^1$ E-mail: j.p.gauntlett@qmul.ac.uk \\
\noindent $^2$ E-mail: j.b.gutowski@qmul.ac.uk}
\end{titlepage}

\sect{Introduction}

There has been recent progress in classifying supersymmetric
bosonic solutions in supergravity theories
\cite{gaunguthull,Gauntlett:2002fz}
(for older work using techniques specific to four dimensions,
see \cite{tod}).
Such a classification is desirable as it may allow one to find new kinds
of solutions that have been hitherto missed by the usual procedure of
starting with an inspired ansatz. In turn this
could elucidate interesting new phenomena in string/M-theory.
In addition such a classification allows one to precisely characterise
supersymmetric geometries of interest, which is important
when explicit solutions are not available.

The basic strategy is to assume the existence of a Killing spinor,
that is assume a solution preserves at least one supersymmetry,
and then consider the differential forms that can be constructed as
bi-linears from the spinor.  These satisfy a number of algebraic and
differential conditions which can be used to determine the form of the
metric and other bosonic fields. Geometrically, the Killing spinor,
or equivalently the differential forms, defines a preferred $G$-structure
and the differential conditions restrict its intrinsic
torsion\footnote{The utility of using $G$-structures in analysing
supersymmetric solutions of supergravity was discussed earlier
in \cite{Friedrich:2001nh,Gauntlett:2002sc}.}.

The analysis of the most general bosonic
supersymmetric solutions of D=11 supergravity was initiated in
\cite{Gauntlett:2002fz}.
It was shown that the solutions always have a Killing
vector constructed as a bi-linear from the Killing spinor and
that it is either time-like or null.
A detailed analysis was undertaken for the time-like case and
it was shown that an $SU(5)$ structure plays a central role in determining
the local form of the most general bosonic supersymmetric
configuration. A similar analysis for the null case, which has
yet to be carried out, would then complete this
classification of the most general supersymmetric
geometries of D=11 supergravity. A finer classification
would be to carry out a similar analysis assuming that there is
more than one Killing spinor and some indications of how
this might be tackled were discussed in \cite{Gauntlett:2002fz}.
Of course, a fully complete classification of D=11 supersymmetric
geometries would require classifying the explicit form of the solutions
within the various classes, but this seems well beyond reach at present.

While more progress on D=11 or 10 supergravity is possible,
it seems a daunting challenge to carry through the programme of
\cite{Gauntlett:2002fz} in full.
Thus, it is of interest to analyse
simpler supergravity theories. In the cases where the theory
arises via dimensional reduction from a higher dimensional
supergravity theory, the analysis can be viewed as classifying
a restricted class of solutions of the higher dimensional theory.
In \cite{gaunguthull} minimal supergravity in D=5 was analysed, which
arises, for example, as a truncation of the dimensional reduction
of D=11 supergravity on a six torus.
As in D=11 supergravity, the general supersymmetric solutions
of the D=5 theory
have either a timelike or a null Killing vector that is constructed
from the Killing spinor. In the time-like case there is an $SU(2)$
structure. More precisely it was shown that working in a neighbourhood
in which the Killing vector is timelike,
the D=5 geometry is completely determined by a
hyper-K\"ahler base-manifold, orthogonal
to the orbits of the Killing vector, along with a function and
a connection one-form
defined on the base that satisfy a pair of simple differential equations.
A similar analysis for the null case, revealed
that the most general solution was a plane fronted wave
determined by three harmonic functions. Although the null case has an
$\bR^3$-structure, this did not play an important role in the analysis.
In addition, it was shown that the generic solutions for both the
time-like and null case preserve 1/2 supersymmetry, but they can also
be maximally supersymmetric.  A further analysis determined the explicit
form of the most general maximally supersymmetric solutions.

Here we shall analyse minimal gauged supergravity in d=5. This theory arises
as a consistent truncation of
the dimensional reduction of type IIB supergravity on a five-sphere.
The gauged theory has the same field content as the ungauged theory,
and given their similarity it is not
surprising that some of the analysis parallels that of \cite{gaunguthull}.
However, it is interesting that there are some
important differences. Once again there are two classes of
supersymmetric solutions, the time-like class and the null class.
In the time-like case, we show that the base manifold of the D=5
geometry orthogonal to the orbits of the Killing vector is now a K\"ahler
manifold with a $U(2)$ structure, and the solutions generically preserve
1/4 of the supersymmetry.
However, in contrast to the ungauged case,
the whole of the geometry is determined by the base-space
up to an anti-holomorphic function on the base. This formalism
thus provides
a very powerful method for the generation of new solutions.

When the Killing vector is null, we show that the five-dimensional solution
is again fixed up to three functions, as in the ungauged case.
However, unlike the ungauged case, these
functions are no longer harmonic, but rather satisfy more complicated
elliptic differential equations on $\bR^3$. These solutions
generically preserve 1/4 of the supersymmetry.

By examining the integrability conditions for the Killing spinor equation
it is simple to show that $AdS_5$ is the unique
solution preserving all supersymmetry. This is in contrast
to the ungauged case where  there is a rich class of
maximally supersymmetric solutions.

By using this formalism, we construct some new solutions of
five dimensional gauged supergravity.
As in the ungauged case, we find that many of the new solutions
have closed time-like curves.
More specifically, we find a family of solutions corresponding to
deformations of $AdS_5$,
in which the deformation depends on a holomorphic function
on a K\"ahler manifold equipped with the Bergmann metric. In the special
case that the holomorphic function is constant, we find a regular
deformation of $AdS_5$ with, for a range of parameters, no closed
time-like curves. We also
find a 1-parameter family of solutions for which the geometry corresponds
to a certain double analytic continuation of the
coset space $T^{pq}$.
All of these solutions can be lifted on a five-sphere
to obtain solutions of type IIB theory
\cite{Chamblin:1999tk,Cvetic:1999xp}.

The plan of this paper is as follows; in section 2 we examine the structure
of the minimal five dimensional gauged supergravity, and describe the
algebraic and differential constraints which bilinears constructed out
of the Killing spinor must satisfy. In section 3, we present a
classification of the solutions when the Killing vector constructed from
the Killing spinor is timelike. We show how the solutions are completely
fixed up to an arbitrary K\"ahler 4-manifold together with an
anti-holomorphic function, and we present some new solutions. In section
4, we examine solutions for which the Killing vector is null; again, we
find a simple prescription for constructing solutions in this
case. In section 5 we investigate maximally supersymmetric solutions.
In section 6 we present our conclusions.

\sect{D=5 Gauged supergravity}

\label{sec:basics}

The bosonic action for minimal gauged supergravity in five dimensions is
\cite{Gunaydin:1983bi}
\be \label{eqn:action}
 S = \frac{1}{4\pi G} \int \left( -\frac{1}{4} ({}^5 R-\c^2) * 1 -
\frac{1}{2} {}F
 \wedge *{}F - \frac{2}{3\sqrt{3}} {}F \wedge {}F \wedge A \right),
\ee
where $F=dA$ is a $U(1)$ field strength and $\c \neq 0$ is a real constant.
We will adopt the same conventions as \cite{gaunguthull}, including
a mostly minus signature for the metric.
The bosonic equations of motion are
\bea\label{eqofmot}
{}^5 R_{\alpha\beta}+2{}F_{\alpha\gamma}{}F_\beta{}^\gamma
-\frac{1}{3}g_{\alpha\beta}{}(F^2+\c^2)&=&0\nn d*{}F +
\frac{2}{\sqrt{3}} {}F \wedge {}F&=0&
\eea
where ${}F^2\equiv {}F_{\alpha\beta}{}F^{\alpha\beta}$. A bosonic
solution to the equations of motion is supersymmetric if it admits
a  super-covariantly constant spinor obeying
\be\label{eqn:kspin}
 \left[ D_\alpha + \frac{1}{4\sqrt{3}}  \left(
 \gamma_\alpha{}^{\beta\gamma} - 4 \delta^{\beta}_{\alpha}\gamma^{\gamma}
  \right){}F_{\beta \gamma} \right] \epsilon^a - \c \epsilon^{ab}
\big( {1 \over 4 \sqrt{3}}\gamma_\alpha -{1 \over 2}A_\alpha \big)
\epsilon^b = 0
\ee
where $\epsilon^a$ is a symplectic Majorana spinor. We shall call such
spinors Killing spinors. Our
strategy for determining the most general bosonic supersymmetric
solutions\footnote{Note that there are spacetimes admitting a Killing spinor
that do not satisfy the equations of motion.
These can be viewed as solutions of the field equations with
additional sources, and
supersymmetry imposes conditions on these sources. It is straightforward
to determine the conditions, but for
simplicity of presentation, we will restrict
ourselves to solutions of the field equations without sources.}
is to analyse the differential forms that can be
constructed from commuting Killing spinors. We first investigate algebraic
properties of these forms, and then their differential properties.

{}From a single commuting spinor $\epsilon^a$ we can construct a scalar $f$,
a 1-form $V$ and three 2-forms $\Phi^{ab} \equiv \Phi^{(ab)}$: \be
 f \epsilon^{ab} = \bar{\epsilon}^a \epsilon^b,
\ee \be
 V_\alpha \epsilon^{ab} = \bar{\epsilon}^a \gamma_\alpha \epsilon^b,
\ee \be
 \Phi^{ab}_{\alpha \beta} = \bar{\epsilon}^a \gamma_{\alpha \beta}
 \epsilon^b,
\ee $f$ and $V$ are real, but $\Phi^{11}$ and $\Phi^{22}$ are
complex conjugate and $\Phi^{12}$ is imaginary. It is   useful
to work with three real two-forms defined by \be
 \Phi^{(11)} = X^{(1)} + i X^{(2)}, \qquad
 \Phi^{(22)} = X^{(1)} - i X^{(2)}, \qquad
 \Phi^{(12)} = - i X^{(3)}.
\ee
It will be useful to record some of these identities which can be
obtained from various {}Fierz identities.

We first note that \be
 V_{\alpha} V^{\alpha} = f^2
\ee which implies that $V$ is timelike, null or zero. The final
possibility can be eliminated using the arguments in
\cite{gaunguthull,Reall:2002bh}.
Now $f$ either vanishes everywhere or
it is non-vanishing at a point $p$. In the former ``null case'', the
Killing vector $V$ is a globally defined null Killing vector. In the
latter ``time-like case'' there is a neighbourhood of $p$ in which $f$
is non-vanishing and for which $V$ is time-like. We will
work in such a neighbourhood for this case, and then find the full
solution by analytic continuation. In later sections we will analyse the
time-like and null cases separately.

We also have \be \label{eqn:XwedgeX}
 X^{(i)} \wedge X^{(j)} = -2\delta_{ij} f * V,
\ee \be \label{eqn:VdotX}
 i_V X^{(i)} = 0,
\ee \be \label{eqn:VstarX}
 i_V * X^{(i)} = - f X^{(i)},
\ee \be \label{eqn:XcontX}
 X^{(i)}_{\gamma \alpha} X^{(j) \gamma}{}_{\beta} = \delta_{ij} \left(
 f^2 \eta_{\alpha\beta} - V_{\alpha} V_{\beta} \right) + \epsilon_{ijk} f
 X^{(k)}_{\alpha\beta}
\ee where $\epsilon_{123} = +1$ and, for a vector $Y$ and $p$-form $A$,
 $(i_Y A)_{\alpha_1 \ldots \alpha_{p-1}} \equiv Y^{\beta} A_{\beta \alpha_1
\ldots \alpha_{p-1}}$. {}Finally, it is useful to record
\be \label{eqn:Vproj}
 V_{\alpha} \gamma^\alpha \epsilon^a = f \epsilon^a \ ,
\ee and \be \label{eqn:Phiproj}
 \Phi^{ab}_{\alpha \beta} \gamma^{\alpha \beta} \epsilon^c = 8f
\epsilon^{c(a} \epsilon^{b)} \ . \ee

We now turn to the differential conditions that can be obtained by
assuming that $\epsilon$ is a Killing spinor. We differentiate
$f$, $V$, $\Phi$ in turn and use ({\ref{eqn:kspin}}). Starting with $f$ we
find \be \label{eqn:df}
 df = -\frac{2}{\sqrt{3}} i_V {}F \ .
\ee
Taking the exterior derivative and using the
Bianchi identity for ${}F$ then gives \be \label{eqn:lie{}F}
 {\cal L}_V {}F = 0 \ ,
\ee where ${\cal L}$ denotes the Lie derivative. Next,
differentiating $V$ gives \be\label{deevee}
 D_\alpha V_\beta = \frac{2}{\sqrt{3}} {}F_{\alpha \beta} f + \frac{1}{2
 \sqrt{3}} \epsilon_{\alpha \beta \gamma \delta \epsilon} {}F^{\gamma
 \delta} V^{\epsilon}+{\c \over 2 \sqrt{3}}(X^1){}_{\alpha \beta} \ ,
\ee which implies $ D_{(\alpha} V_{\beta)} = 0$ and hence $V$ is a
Killing vector.
Combining this with
\p{eqn:lie{}F} implies that $V$ is the generator of a symmetry of
the full solution $(g,{}F)$. Note that \p{deevee} implies
\be
\label{eqn:dV}
 dV = \frac{4}{\sqrt{3}} f {}F + \frac{2}{\sqrt{3}} * ({}F \wedge V)
+{\c \over \sqrt{3}} X^1 \ .
\ee
{}Finally, differentiating $X^{(i)}$ gives
\bea \label{eqn:dPhi}
 D_\alpha X^{(i)}_{\beta\gamma} &=& -\frac{1}{\sqrt{3}} \left[ 2
 {}F_{\alpha}{}^{\delta} \left( *X^{(i)} \right)_{\delta\beta\gamma} -2
 {}F_{[\beta}{}^{\delta} \left( *X^{(i)} \right)_{\gamma] \alpha
 \delta} + \eta_{\alpha [\beta} {}F^{\delta \epsilon} \left( *X^{(i)}
 \right)_{\gamma] \delta\epsilon} \right]
\nn
&-& {\c \over \sqrt{3}}\delta^{1i} \eta_{\alpha [\beta}V_{\gamma]}
+\c \epsilon^{1ij} \big[A_\alpha X^{(j)}{}_{\beta \gamma}
+{1 \over 2 \sqrt{3}} (*X^{(j)}){}_{\alpha \beta \gamma} \big] \ .
\eea
Note that ({\ref{eqn:dPhi}}) implies that
\be
\label{eqn:cclos}
dX^{(i)} = \c \epsilon^{1ij} \big(A \wedge X^{(j)}+{\sqrt{3} \over 2}
*X^{(j)} \big)
\ee
so $dX^{(1)}=0$ but $X^{(2)}$ and $X^{(3)}$ are not closed.
In particular, this implies that
\be
\label{eqn:lied}
\cL_V X^{(i)} = \c \epsilon^{1ij} (i_V A -{\sqrt{3} \over 2}f) X^{(j)} \ .
\ee

It is useful to consider the effect of gauge transformations $A \rightarrow
A+d\Lambda$. In particular, the Killing spinor equation is left
invariant under the transformation
\bea
\label{eqn:kspmap}
\epsilon^1 &\rightarrow& \cos ({\c \Lambda \over 2}) \epsilon^1
-  \sin ({\c \Lambda \over 2}) \epsilon^2
\nn
\epsilon^2 &\rightarrow& \cos ({\c \Lambda \over 2}) \epsilon^2
+  \sin ({\c \Lambda \over 2}) \epsilon^1 \ .
\eea
Under these transformations, $f \rightarrow f$, $V \rightarrow V$
and $X^1 \rightarrow X^1$, but $X^2 +i X^3 \rightarrow e^{-i \c \Lambda}
(X^2+i X^3)$. We shall choose to work in a gauge in which
\be\label{gaugechoice}
i_V A ={\sqrt{3} \over 2}f
\ee
and so $\cL_V A=0$ and also $\cL_V X^{(i)}=0$.

To make further progress we will examine separately the case in which
the  Killing vector is time-like and the case in which it is null in the
two following sections.

\sect{The timelike case}

\label{sec:timelike}

\subsection{The general solution}

In this section we shall consider solutions in a neighbourhood
in which $f$ is non-zero and hence $V$ is a timelike Killing vector field.
Equation \p{eqn:XcontX} implies that the 2-forms $X^{(i)}$ are all
non-vanishing. Introduce coordinates such that $V =
\partial/\partial t$. The metric can then be written locally as
\be
\label{eqn:metric} ds^2=f^2(dt+\omega)^2-f^{-1}h_{mn}dx^m dx^n \ee
where we have assumed, essentially with no loss of generality, $f>0$;
$f$, $\omega$ and $h$ depend only on $x^m$ and not on $t$.
The metric $f^{-1} h_{mn}$ is obtained by
projecting the full metric perpendicular to the orbits of $V$. The
manifold so defined will be referred to as the base space $B$.

Define
\be \label{eqn:e0def}
 e^0 = f (dt+\omega)
\ee
and if $\eta$ defines a positive orientation on $B$ then we
define $e^0\wedge \eta$ to define a positive orientation for the D=5 metric.
The two form $d\omega$ only has components tangent to the base
space and can therefore be split into self-dual and anti-self-dual
parts with respect to the metric $h_{mn}$:
\be
\label{eqn:rsp}
f d\omega=G^{+}+G^{-}
\ee
where the factor of $f$ is included for convenience.

Equation \p{eqn:VdotX} implies that the $2$-forms $X^{(i)}$ can be
regarded as $2$-forms on the base space and Equation
\p{eqn:VstarX} implies that they are anti-self-dual: \be
 *_4 X^{(i)} = - X^{(i)},
\ee where $*_4$ denotes the Hodge
dual associated with the metric
$h_{mn}$. Equation \p{eqn:XcontX} can be written \be
\label{eqn:quat}
 X^{(i)}{}_m{}^p X^{(j)}{}_p{}^n = - \delta^{ij} \delta_m{}^n
 + \epsilon_{ijk} X^{(k)}{}_m{}^n
\ee where indices $m,n, \ldots$ have been raised with $h^{mn}$,
the inverse of $h_{mn}$. This equation shows that the $X^{(i)}$'s
satisfy the algebra of imaginary unit quaternions.

To proceed, we use ({\ref{eqn:df}}) and ({\ref{eqn:dV}}) to solve for the
gauge field strength $F$. This gives
\be
\label{eqn:Fsol}
F ={\sqrt{3} \over 2} de^0 -{1 \over \sqrt{3}} G^+ -{\c \over 2 f} X^{(1)}
\ .
\ee
It is convenient to write
\be
\label{eqn:hdef}
H = {\sqrt{3} \over 2} de^0 -{1 \over \sqrt{3}} G^+
\ee
so that $F=H  -{\c \over 2 f} X^{(1)}$. Substituting this
into ({\ref{eqn:dPhi}}) we find that
\bea
\label{eqn:simpdfx}
 D_\alpha X^{(i)}_{\beta\gamma} &=& -\frac{1}{\sqrt{3}} \left[ 2
 {}H_{\alpha}{}^{\delta} \left( *X^{(i)} \right)_{\delta\beta\gamma} -2
 {}H_{[\beta}{}^{\delta} \left( *X^{(i)} \right)_{\gamma] \alpha
 \delta} + \eta_{\alpha [\beta} {}H^{\delta \epsilon} \left( *X^{(i)}
 \right)_{\gamma] \delta\epsilon} \right]
\nn
&+& \c \epsilon^{1ij} (A_\alpha -{\sqrt{3} \over 2f}V_\alpha)
X^{(j)}_{\beta \gamma} \ .
\eea
We also find that
\bea
\label{eqn:nothk}
\nabla_m X^{(1)}_{np} &=& 0
\nn
\nabla_m X^{(2)}_{np} &=& P_m X^{(3)}_{np}
\nn
\nabla_m X^{(3)}_{np} &=& -P_m X^{(2)}_{np}
\eea
where $\nabla$ is the Levi-Civita connection on $B$ with respect to $h$
and we have introduced
\be\label{defbig}
P_m=\c \big(A_m- {\sqrt{3} \over 2} f \omega_m \big) \ .
\ee
Recall that $X^{(1)}$ is gauge-invariant.
{}From \p{eqn:quat} and \p{eqn:nothk} we conclude that
the base space is K\"ahler, with K\"ahler form $X^{(1)}$.
Thus the base space has a $U(2)$ structure.

One might be tempted to conclude that the additional
presence of $X^{(2)}$ and $X^{(3)}$ satisfying \p{eqn:quat}
implies that the manifold actually has an $SU(2)$ structure.
However, this is not the case since $X^{(2)}$ and $X^{(3)}$
are not gauge invariant.
To obtain some further insight, note that we can invert ({\ref{eqn:nothk}})
to solve for $P$:
\be
\label{eqn:invrta}
P_m ={1 \over 8 } \big( X^{(3) np} \nabla_m X^{(2)}_{np}-
X^{(2) np} \nabla_m X^{(3)}_{np} \big)
\ee
from which we deduce that
\be\label{defpee}
dP=\Re
\ee
where $\Re$ is the Ricci-form of the base space $B$ defined by
\be
\Re_{mn}={1 \over 2} X^{(1) pq} R_{pq mn}
\ee
and $R_{pqmn}$ denotes the Riemann curvature tensor
of $B$
equipped with metric $h$.
Now on any K\"ahler four-manifold, with anti-self dual  K\"ahler
two-form $X^{(1)}$ and Ricci-form $\Re$, there is always a
section of the canonical bundle, $X^{(2)}+iX^{(3)}$, with
 anti-self-dual two-forms $X^{(2)}, X^{(3)}$, satisfying \p{eqn:quat},
and $(\nabla +iP)(X^{(2)}+iX^{(3)})=0$. But this is equivalent to the
last two equations in \p{eqn:nothk}.
(Note that shifting $P$ by
a gradient of a function on the K\"ahler manifold, shifts
$X^{(2)}+iX^{(3)}$ by a phase, which precisely corresponds to
the time-independent gauge transformations of $X^{(2)}+iX^{(3)}$.)

Thus the content of \p{eqn:quat}, \p{eqn:nothk} and \p{defbig} is simply
that the base $B$ is K\"ahler and that the base determines
$A_m- {\sqrt{3} \over 2} f \omega_m$
(up to a gradient of a time independent function).
In fact, as we now show all of the five-dimensional geometry is determined
in terms of the geometry of the base space $B$, up
to an anti-holomorphic function on the base.
To see this we first substitute \p{defbig} into \p{defpee} to get
\be
\label{eqn:strngecurvident}
-{1 \over \sqrt{3}} G^+{}_{mn} -{\c \over 2 f} X^{(1)}_{mn} =
{1\over \chi}\Re_{mn} \ .
\ee
Upon contracting ({\ref{eqn:strngecurvident}}) with $(X^1)^{mn}$, and
using $\Re_{mn} X^{(1)mn}=R$, we obtain
\be
\label{eqn:ffix}
f = -{2 \c^2 \over R}
\ee
where $R$ is the Ricci scalar curvature of $B$.
In particular, we see $B$ cannot be hyper-K\"ahler, as we must have
$R \neq 0$. Substituting back into
({\ref{eqn:strngecurvident}}) we find that
\be
\label{eqn:gpfix}
G^+{}_{mn} = - {\sqrt{3} \over\c} \big( \Re_{mn}-{1 \over 4}
R X^{(1)}_{mn} \big) \ .
\ee
Now the Bianchi identity $dF=0$ is satisfied since
\be
\label{eqn:timbianc}
d G^+ = {\sqrt{3} \c \over 2 f^2} df \wedge X^1
\ee
which is implied by  ({\ref{eqn:strngecurvident}}).
The gauge field equation implies that
\be
\label{eqn:gfdeq}
\nabla_m \nabla^m f^{-1} = {2 \over 9}(G^+)^{mn} (G^+)_{mn}+
{\c \over 2 \sqrt{3} f} (G^-)_{mn} (X^1)^{mn}
-{2 \c^2 \over 3 f^2} \ .
\ee
If we write
\be
\label{eqn:expandit}
G^- = \lambda^i X^i,
\ee
for some functions $\lambda^i$,
we see that ({\ref{eqn:gfdeq}}) fixes $\lambda^1$ in terms of the
 base space geometry via
\be
\label{eqn:lifix}
\lambda^1 = {{\sqrt 3} \over \c R} \big( {1\over 2}
\nabla^m \nabla_m  R +{2 \over {3}}
\Re_{mn} \Re^{mn} -{1 \over {3}} R^2 \big) \ .
\ee
Next we note that ({\ref{eqn:rsp}}) implies that $R (G^+ + G^-)$ is closed.
Hence, on taking the
exterior derivative and using ({\ref{eqn:nothk}}) we find that
\be
\label{eqn:geodata}
T + \big[ d (R \lambda^2) - R \lambda^3 P \big] \wedge X^2 +
 \big[ d (R \lambda^3) +  R \lambda^2 P \big] \wedge X^3=0
\ee
where
\be
\label{eqn:tdef}
T = {\sqrt{3} \over \c} \left(-d R \wedge \Re +d \big[{1 \over 2}
 \nabla_m \nabla^m R
+{2 \over 3} \Re_{mn} \Re^{mn}-{1 \over 12} R^2 \big] \wedge X^1 \right)
\ee
is determined by the geometry of the base. In particular
$\lambda^2=\lambda^3=0$ is only possible if and only if $T=0$.
On defining
\be
\label{eqn:geodatb}
\Theta_m = (X^2)_m{}^n (*_4 T)_n
\ee
and adopting complex co-ordinates $z^j, {z^{\bar j}}$
on $B$ with respect to $X^1$, ({\ref{eqn:geodata}}) simplifies to
\be
\label{eqn:geodatc}
\Theta_j = -\left(\partial_j -iP_j\right)\big[R(\lambda^2-i \lambda^3)\big]
\ee
which fixes $\lambda^2 -i \lambda^3$ up to an
arbitrary anti-holomorphic function.
In summary, we have determined $f$ and $G^\pm$ in terms of the K\"ahler
base up to an anti-holomorphic function; then,
up to a time independent gradient, $\omega$ is determined by
({\ref{eqn:rsp}}), and then $A_m$ by $P_m$. This state of affairs
should be contrasted with the ungauged case \cite{gaunguthull},
where $f$ and $\omega$
satisfied a pair of differential equations on a hyper-K\"ahler base.

We remark that there are no solutions for which $V$ is hyper-surface
orthogonal; in other words there are no solutions with $d \omega =0$.
To see this note that if $d \omega =0$ then $G^+=G^-=0$, and from
({\ref{eqn:timbianc}}) we find that $df=0$. On substituting this
into ({\ref{eqn:gfdeq}}) and using $G^+=G^-=0$ we obtain a contradiction.
This would seem problematic, as it is known
that many of the known solutions such as $AdS_5$ and certain
types of nakedly singular black hole solutions can be written in
co-ordinates in which the solution is static with respect to some
time-like killing vector. This apparent contradiction is resolved by
noting that this time-like killing vector is not
the killing vector constructed from the Killing spinor. Hence, it is
clear that the co-ordinates  which arise naturally from the construction
described here are not in fact the co-ordinates in which
the known solutions can be written in a static form. This is, however, a
minor inconvenience in recovering the known solutions, since
as we shall see, the two co-ordinate systems are typically related
by rather simple co-ordinate transformations. Moreover,
it is clear that the formalism described above is particularly
useful in generating new solutions.

We have obtained all of the constraints on the bosonic quantities $f$, $V$
and $X^{(i)}$ imposed by the Killing spinor equations and the equations
of motion. It remains to check whether, conversely, the geometry we have
found always admits Killing spinors. We shall impose the constraint
\be
\label{eqn:smpkspc}
\epsilon^a = {1 \over 4} \epsilon^{ab} (X^1)_{AB} \Gamma^{AB} \epsilon^b
\ee
where $e^A$ denotes a vielbein adapted for the K\"ahler base space and
$\Gamma_A = i \gamma_A$ satisfy
\be
\label{eqn:gammat}
\Gamma_m \Gamma_n + \Gamma_n \Gamma_m =2 h_{mn} \ .
\ee
The constraint ({\ref{eqn:smpkspc}}) reduces the number of degrees of
freedom in the Killing spinor from 8 to 2, hence 1/4 of the supersymmetry
is preserved. Note also that ({\ref{eqn:smpkspc}}) implies that $\gamma^0
\epsilon^a = \epsilon^a$.
Using this constraint we note that
the Killing spinor equation can be rewritten in terms of $H$ as
\bea
\label{eqn:ssmpkspa}
\big[D_\alpha +{1 \over 4 \sqrt{3}}\big(\gamma_\alpha{}^{\beta \lambda}
- \delta_\alpha^\beta \gamma^\lambda \big)H_{\beta \lambda}
+{\sqrt{3} \c \over 4 f}(X^1)_{\alpha \lambda} \gamma^\lambda \big]
\epsilon^a-{\c \over 2} \epsilon^{ab} \big({\sqrt{3} \over 2}
\gamma_\alpha-A_\alpha\big)\epsilon^b =0 \ .
\eea
Then, using \p{gaugechoice},
the $t$ component of ({\ref{eqn:ssmpkspa}}) requires
that ${\partial \epsilon^a
\over \partial t}=0$, so that $\epsilon^a$ depends only on the $x^m$.
Next we consider the $m$ component of  ({\ref{eqn:ssmpkspa}}); it is
convenient to re-scale
\be
\label{eqn:rescksp}
\epsilon^a = f^{1 \over 2} \eta^a \ ,
\ee
and we then obtain
\be
\label{eqn:smpkspb}
\nabla_m \eta^a +{1 \over 2} P_m
\epsilon^{ab} \eta^b =0 \ .
\ee
Since ({\ref{eqn:smpkspb}}) always admits
two linearly independent solutions on a
K\"ahler manifold given the projections ({\ref{eqn:smpkspc}}),
(see e.g. \cite{popeqg}), we have shown that the geometry does indeed admit
Killing spinors; and generically we find that the timelike
solutions preserve at least 1/4 of the supersymmetry{\footnote{
Note that one can use the spinorial construction of
$X^{(2)}, X^{(3)}$ to show that a K\"ahler manifold always satisfies
\p{eqn:nothk}.}.

\subsection{Some Examples}

Using the techniques described in the previous section, it is possible to
construct gauged supergravity solutions with timelike $V$. In the following,
we shall denote an orthonormal basis of the K\"ahler base space $B$ by
$\{e^1 , e^2, e^3 , e^4 \}$ and take $e^1 \wedge e^2 \wedge e^3 \wedge e^4$
to define a positive orientation with
\bea
\label{eqn:xexprs}
X^1 &=& e^1 \wedge e^2 - e^3 \wedge e^4
\nn
X^2 &=& e^1 \wedge e^3 + e^2 \wedge e^4
\nn
X^3 &=& e^1 \wedge e^4 - e^2 \wedge e^3 \ .
\eea

\subsubsection{Bergmann base space and deformations of $AdS_5$}

The simplest class of examples are those for which the base space $B$ is
 Einstein. From ({\ref{eqn:strngecurvident}}) we see that this is equivalent
to $G^+=0$. Moreover, if $G^+=0$ then from ({\ref{eqn:timbianc}}) we obtain
 $df=0$, and without loss of generality we set $f=1$, and so $R = -2 \c^2$
 and $\Re = -{\c^2 \over 2}X^1$. Hence, from ({\ref{eqn:lifix}}) we find
 $\lambda^1 = {\c \over \sqrt{3}}$ and we note that $\Theta=0$.
Hence, locally ({\ref{eqn:geodatc}}) can be written as
\be
\label{eqn:einy}
\partial_j (\lambda^2-i \lambda^3)+{\c^2 \over 4} \partial_j K
 (\lambda^2-i \lambda^3)=0
\ee
where $K$ is the K\"ahler potential of $B$, so
\be
\label{eqn:eingmin}
\lambda^2-i \lambda^3=e^{-{\c^2 K \over 4}} {\cal{F}} ({\bar{z}})
\ee
where $ {\cal{F}} ({\bar{z}})$ is an anti-holomorphic function. Note that
the field strength takes the simple form
\be
F=\frac{\sqrt 3}{2}(\lambda^2 X^2 +\lambda^3 X^3) \ .
\ee

A simple Einstein base is obtained by taking the base metric to
be given by the Bergmann metric
\be
\label{eqn:einsta}
ds^2 = dr^2+ {3 \over \c^2} \sinh^2 ({\c r \over 2 \sqrt{3}})
\big((\sigma^L_1)^2+(\sigma^L_2)^2\big)
+{3 \over \c^2}  \sinh^2 ({\c r \over 2 \sqrt{3}})  \cosh^2
({\c r \over 2 \sqrt{3}}) (\sigma^L_3)^2
\ee
where $\sigma_L^i$ are right invariant one-forms on the three-sphere and
we use the same Euler angles and notation as in \cite{gaunguthull}.
The $SU(2)$ structure is given by \p{eqn:xexprs} if we choose the
orthonormal basis
\be
\label{eqn:onbasisa}
e^1 = dr \ , \ e^2 =  {\sqrt{3} \over \c}  \sinh
({\c r \over 2 \sqrt{3}})  \cosh
({\c r \over 2 \sqrt{3}}) \sigma^L_3 \ , \ e^3 =  {\sqrt{3} \over \c} \sinh
({\c r \over 2 \sqrt{3}}) \sigma^L_1
\ ,  \ e^4 = {\sqrt{3} \over \c} \sinh ({\c r \over 2 \sqrt{3}})
\sigma^L_2\ee
More explicitly we have
\bea
\label{eqn:basedetails}
X^1 &=& {3 \over \c^2} d \big[ \sinh^2 ({\c r \over 2 \sqrt{3}})
\sigma^L_3 \big
],\nn
X^2 &=& {3 \over \c^2} \cosh^3 ({\c r \over 2 \sqrt{3}}) d \big[
\tanh^2 ({\c r \over 2 \sqrt{3}}) \sigma^L_1 \big],
\nn
X^3 &=& {3 \over \c^2} \cosh^3 ({\c r \over 2 \sqrt{3}}) d \big[
\tanh^2 ({\c r \over 2 \sqrt{3}}) \sigma^L_2 \big]\ ,
\eea
and $P = -{3 \over 2} \sinh^2 ({\c r \over 2 \sqrt{3}}) \sigma^L_3$.

{}For solutions with ${\cal{F}}=0$ (and so $\lambda^2=\lambda^3=0$ and $F=0$)
we find $\omega={\sqrt{3} \over \c}  \sinh^2 ({\c r \over 2 \sqrt{3}})
\sigma^L_3 $. The five dimensional
geometry can be written, after shifting the Euler angle
$\phi\to \phi+\c/{\sqrt 3}t$, as
\be
\label{eqn:adsv}
ds^2 = \cosh^2 ({\c r \over 2 \sqrt{3}}) dt^2 -dr^2 - {12 \over \c^2}
 \sinh^2 ({\c r \over 2 \sqrt{3}}) d \Omega_3{}^2
\ee
which is the simply the metric of $AdS_5$ with radius $2{\sqrt 3}/\chi$.

In order to construct new solutions with $F \neq 0$ we
exploit the fact that the  K\"ahler potential is well known
in complex co-ordinates (see e.g. \cite{rbpstrom}).
In particular if we introduce
the complex coordinates
\bea
\label{eqn:coordtrans}
z^1 &=& \tanh ({\c r \over 2 \sqrt{3}}) \cos ({\theta \over 2})
 e^{{i \over 2}(\phi+\psi)}
\nn
z^2 &=& \tanh ({\c r \over 2 \sqrt{3}}) \sin ({\theta \over 2})
 e^{{i \over 2}(\phi-\psi)}
\eea
the K\"ahler potential is
\be
\label{eqn:bergkp}
K = -{6 \over \c^2} \log (1-|z^1|^2-|z^2|^2) \ .
\ee
Thus in the real coordinates,
$K={12 \over \c^2} \log \cosh ({\c r \over 2 \sqrt{3}}) $
and hence
$\lambda^2-i \lambda^3 = \cosh^{-3} ({\c r \over 2 \sqrt{3}})
{\cal{F}} ({\bar{z}})$.
If we write ${\cal{F}}\equiv \cF_1-i\cF_2$ then we find
\be
\label{eqn:omsol}
d\omega = d\left[{\sqrt{3} \over \c}
\sinh^2 ({\c r \over 2 \sqrt{3}}) \sigma^L_3\right]
+\cF_1d\left[{3\over\c^2} \tanh^2 ({\c r \over 2 \sqrt{3}}) \sigma^L_1
\right]
+\cF_2d\left[{3\over\c^2} \tanh^2 ({\c r \over 2 \sqrt{3}}) \sigma^L_2
\right] \ .
\ee
It would be interesting to explore these deformation of $AdS_5$ in more
detail. Let us just note here that if we consider the special case when
$\cF_i$ are constant, it is trivial to find the explicit form of $\omega$.
Interestingly, this case seems to be a completely regular deformation of
$AdS_5$. Moreover, by considering
the norm of the left vector fields $\xi^L_1$ and
$\xi^L_2$, we find that there are closed
time-like curves, for sufficiently small $r$, when
$\cF_i^2 > {4 \over 3} \c^2$ and they appear to be absent
otherwise.

\subsubsection{Base space is a product of two-manifolds}

Let us now consider some examples in which the base manifold is a
product $B=M_2 \times N_2$ where
$M_2$, $N_2$ are two 2-manifolds. When the base space is itself
not Einstein, then these solutions have $G^+ \neq 0$.
In the first case, we take $B= H^2 \times H^2$ equipped with metric
\be
\label{eqn:hiihiimet}
ds^2 = (dr^2+ \sinh^2 r d \theta^2)+\beta^2(d \rho^2+ \sinh^2 \rho d \phi^2)
\ee
for $\beta$ constant. Note that  setting the radius of the first factor
to one, as we have done, does not in fact result in any loss of generality
in the resulting five dimensional geometries. Clearly this base is
Einstein iff $\beta^2=1$.
We take the orthonormal basis to be
\be
\label{eqn:onbasis}
e^1 = dr \ , \quad e^2= \sinh r d \theta \ , \quad e^3 =
\beta d \rho \ , \quad e^4 = \beta \sinh \rho d \phi \ .
\ee
It is straightforward to show that for this solution
\bea
\label{eqn:solutdatb}
P &=&  - \coth r e^2 + \beta^{-1} \coth \rho e^4
\nn
f &=&  {\c^2 \beta^2 \over 1+\beta^2}
\nn
G^+ &=& {\sqrt{3} (\beta^2 - 1) \over 2 \c \beta^2}
 (e^1 \wedge e^2+ e^3 \wedge e^4)
\nn
\lambda^1 &=& {4 \over \sqrt{3} \c (1+\beta^2)}
\nn
\Theta &=& 0 \ .
\eea
Since $\Theta=0$ we can set $\lambda^2=\lambda^3=0$, which we do
for simplicity. We then find that
\be
\label{eqn:omsolb}
f \omega = {1 \over 2 \sqrt{3} \c (1+ \beta^2)} \big[ \beta^{-2}
(3 \beta^2-1)(\beta^2+3)
\cosh r d \theta +(\beta^2-3)(3 \beta^2+1) \cosh \rho d \phi \big]
\ee
where
\be
\label{eqn:Fhypersol}
F = {(\beta^2-1) \over 4 \c (1+ \beta^2)} \big[ \beta^{-2} (3-\beta^2)
\sinh r dr \wedge d \theta
+(3 \beta^2-1) \sinh \rho d \rho \wedge d \phi \big] \ .
\ee
After re-scaling $t= {(1+\beta^2) \over \c^2 \beta^2} t'$
we find
\bea
\label{eqn:simplergeom}
ds^2 &=& \big[ dt' +{1 \over 2 \sqrt{3} \c (1+ \beta^2)}
\big\{ \beta^{-2} (3 \beta^2-1)(\beta^2+3)
\cosh r d \theta +(\beta^2-3)(3 \beta^2+1) \cosh \rho d \phi \big\} \big]^2
\nn
&-& {(1+\beta^2) \over \c^2} \big[{1 \over \beta^2} (dr^2+ \sinh^2 r
d \theta^2)+(d \rho^2+ \sinh^2 \rho d \phi^2) \big] \ .
\eea
{}From these expressions we observe that the solution remains unchanged
(up to a co-ordinate transformation)
under the operation $\beta \rightarrow {1 \over \beta}$.

There are two special cases to consider. Firstly, when $\beta=1$ we
obtain the geometry
\be
\label{eqn:lamone}
ds^2 = \big[d{\hat{t}} +{2 \over \sqrt{3} \c}(\cosh r d \theta -
\cosh \rho d \phi) \big]^2
-{2 \over \c^2} \big[ dr^2+ \sinh^2 r d \theta^2 +  d \rho^2+
\sinh^2 \rho d \phi^2 \big]
\ee
with $F=0$. This an Einstein metric, admitting a Killing spinor (it is not
maximally symmetric and so it is not $AdS_5$).
Secondly, if we take $\beta= {1 \over \sqrt{3}}$
we obtain
\be
\label{eqn:lamonetre}
ds^2 = \big[d {\hat{t}} -{2 \over \sqrt{3} \c} \cosh \rho d \phi \big]^2
-{4 \over 3 \c^2}
\big[ 3 (dr^2+ \sinh^2 r d \theta^2) + (d \rho^2+ \sinh^2 \rho d \phi^2)
\big]
\ee
with $F=- \c^{-1} \sinh r dr \wedge d \theta$. This is the metric of
$AdS_3 \times H^2$,
and we recover the near horizon limit of the supersymmetric
black string solution with hyperbolic transverse space \cite{sabrahyp}.

Thus our general solution, \p{eqn:simplergeom}, \p{eqn:Fhypersol},
is a one parameter family of supersymmetric solutions interpolating
between the Einstein metric \p{eqn:lamone} and $AdS_3 \times H^2$.
Note that for the entire family of solutions
there are closed timelike curves in the neighbourhood of $r=0$ or
$\rho=0$ parallel to
${\partial \over \partial \theta}$ or ${\partial \over \partial \phi}$
respectively. Of course we know that the closed timelike curves can be
eliminated for $AdS_3 \times H^2$ by going to the covering space, and
it would be interesting to know if this happens for the entire family of
solutions. {}Finally, we note that if we perform a double analytic
continuation $\theta \to i\theta$, $\phi\to i\phi$, and periodically
identify the time co-ordinate, we see from the discussion in, for example,
\cite{gubconf}, that the metric is that on the coset space
$T^{p,q}=SU(2) \times SU(2) /U(1)_{p,q}$
with squashing parameterized by $\beta$ and $p$, $q$ are related via
\be
\label{eqn:pqrel}
\beta^{-2} (3 \beta^2-1)(\beta^2+3)p-(\beta^2-3)(3 \beta^2+1)q=0 \ .
\ee

The second class of solutions is obtained when we take the base space to
be $B=H^2 \times S^2$. In fact this solution
can be obtained from the expressions given above on mapping
$\rho \rightarrow i \rho$
(and restricting $0<\rho<\pi$) and
$\beta \rightarrow -i \beta$. We thus
find that the solution, with $\lambda^2=\lambda^3=0$, can be written
\bea
\label{stwohtwo}
ds^2 &=& \big[ dt' +{1 \over 2 \sqrt{3} \c (1-\beta^2)}
\big\{ \beta^{-2} (3 \beta^2+1)(-\beta^2+3)
\cosh r d \theta -
(\beta^2+3)(-3 \beta^2+1) \cos \rho d \phi \big\} \big]^2
\nn
&-& {(\beta^2-1) \over \c^2} \big[{1 \over \beta^2} (dr^2+ \sinh^2 r d
\theta^2)+(d \rho^2+ \sin^2 \rho d \phi^2) \big]\nn
F &=& {(\beta^2+1) \over 4 \c (1-\beta^2)}
\big[ \beta^{-2} (3+\beta^2) \sinh r dr \wedge d \theta
-(3 \beta^2+1) \sin \rho d \rho \wedge d \phi \big] \ .
\eea
In contrast to the previous solution, it is clear that
we must have $\beta>1$.
Thus, in this case, it is not possible to choose $\beta$ in such a way as to
obtain an Einstein metric. By considering the norm of the vector
$\partial_\phi$ we see that the solutions have closed time-like curves.
It is also interesting to note that for the special
solution $\beta^2=3$ the metric becomes a direct product of a three-space
with $H^2$.

{}For a final example of a solution with product base space, we take the
base to be $B=M_2 \times \bR^2$ with metric
\be
\label{eqn:prodmetiii}
ds^2 = {1 \over r^2 (\alpha r^2+ {\beta \over r^4})} dr^2+r^4
(\alpha r^2+ {\beta \over r^4})
dz^2 + dx^2+dy^2
\ee
for positive constants $\alpha$, $\beta$; and we take an orthonormal basis
\be
\label{eqn:oobasinvol}
e^1 = {1 \over r \sqrt{\alpha r^2+ {\beta \over r^4}}} dr
\ , \ e^2=r^2 \sqrt{\alpha r^2+ {\beta \over r^4}} dz \ , \ e^3 =dx \ ,
\ e^4 =dy \ .
\ee
This solution has
\bea
\label{eqn:solutdatc}
P &=&  - 3 \alpha r^4 dz
\nn
f &=&  {\c^2 \over 12 \alpha r^2}
\nn
G^+ &=& {6 \sqrt{3} \alpha r^2 \over \c}(e^1 \wedge e^2+ e^3 \wedge e^4)
\nn
\lambda^1 &=& {6 \sqrt{3} \alpha r^2 \over \c}
\nn
\Theta &=& 0 \ .
\eea
{}For simplicity we set $\lambda^2=\lambda^3=0$ and obtain
\be
\label{eqn:wwsols}
\omega = {24 \alpha^2 \sqrt{3} r^6 \over \c^3} dz \ .
\ee
The solution has $F= -{\sqrt{3} \over 2} dt \wedge df$ and setting
$z={z' \over \sqrt{\beta}} +{\c^3 t \over 24 \sqrt{3} \alpha \beta}$ the
metric simplifies to
\be
\label{eqn:infvolmet}
ds^2 = {\c^4 \over 144 \alpha^2 \beta} (\alpha r^2+ {\beta \over r^4})dt^2
- {12 \alpha \over \c^2} (\alpha r^2+ {\beta \over r^4})^{-1} dr^2 -
{12 \alpha r^2 \over \c^2} ds^2 (\bR^3) \ .
\ee
This metric is a supersymmetric ``topological black hole''
\cite{Chamblin:1999tk}
and it can be obtained from taking the infinite volume limit of
the nakedly-singular supersymmetric ``black hole'' solution to be discussed
next.

\subsubsection{Black Hole Solutions}

In order to obtain black hole solutions we shall set the metric on
 the base manifold to be
\be
\label{eqn:basebh}
ds^2 = H^{-2} dr^2 +{r^2 \over 4}H^2 (\sigma^L_3)^2+{r^2 \over 4}
[(\sigma^L_1)^2+(\sigma^L_2)^2]
\ee
with orthonormal basis
\be
\label{eqn:bhortho}
e^1 = H^{-1} dr \ , \quad e^2 ={rH \over 2} \sigma^L_3 \ , \quad e^3 =
{r \over 2} \sigma^L_1 \ , \quad e^4 = {r \over 2} \sigma^L_2
\ee
and we set
\be
\label{eqn:hhsol}
H = \sqrt{1+{\c^2 \over 12} r^2(1+{\mu \over r^2})^3}\ .
\ee
With this choice of $H$,
$\Theta=0$. Once again, this allows us to set
$\lambda^2=\lambda^3=0$ for simplicity. Moreover,
\bea
\label{eqn:morebhinfo}
P &=& -{\c^2 \over 8 r^2}(r^2+\mu)^2 \sigma^L_3
\nn
\omega &=& {\c \over 4 \sqrt{3} r^4} (r^2+ \mu)^3 \sigma^L_3
\nn
f &=&  (1+{\mu \over r^2})^{-1}
\nn
\lambda^1 &=& {\c \over 2 \sqrt{3} r^4} (r^2+\mu)(2r^2- \mu)
\nn
G^+ &=& -{\sqrt{3} \c \mu \over 2 r^4} (r^2+\mu) (e^1 \wedge e^2+ e^3
\wedge e^4) \ .
\eea
On setting $\phi = \phi' + {\c \over \sqrt{3}} t$ the spacetime
geometry simplifies to
\be
\label{eqn:nsingbh}
ds^2 = f^2 (1+ {\c^2 \over 12} r^2 f^{-3})dt^2- f^{-1}
\big[(1+ {\c^2 \over 12} r^2 f^{-3})^{-1} dr^2 + r^2 d \Omega_3^2 \big]
\ee
with $F=-{\sqrt{3} \over 2} dt \wedge df$, where $d \Omega_3^2$
denotes the metric on $S^3$. These are the supersymmetric
black holes, with naked singularities,
first constructed in \cite{londi} (to get the same
coordinates shift $r^2=R^2-\mu$).
On taking the ``infinite volume'' limit, in which the 3-sphere blows-up to
$\bR^3$, we recover, up to a co-ordinate transformation the metric
({\ref{eqn:infvolmet}}) \cite{Chamblin:1999tk}.
Note that on holding $\mu$ constant and
letting $\c \rightarrow 0$, we obtain,
as expected, the electrically charged static black hole solution
of the ungauged theory.

We remark that all of these timelike solutions have $\Theta =0$, which
is a strong restriction on the base. It
would appear therefore that there is a rich structure of
new solutions for which $\Theta \neq 0$.
It would be interesting to see if
the rotating black hole solutions examined in \cite{sabbh}
lie within this class.

\sect{The null case}

\label{sec:null}

\subsection{The general solution}

In this section we shall find all solutions of minimal gauged $N=1$, $D=5$
supergravity for which the function $f$ introduced in section 2
vanishes everywhere.

{}From \p{eqn:dV} it can
be seen that $V$ satisfies $V\wedge dV=0$ and is therefore
hypersurface-orthogonal. Hence there exist functions $u$ and $H$
such that \be \label{eqn:udef}
 V = H^{-1} du \ .
\ee
A second consequence of \p{deevee} is
\be
 V \cdot D V = 0 \ ,
\ee
so $V$ is tangent to affinely parameterized geodesics in the
surfaces of constant $u$. One can choose coordinates $(u,v,y^m)$,
$m=1,2,3$, such that $v$ is the affine parameter along these
geodesics, and hence
\be V = \frac{\partial}{\partial v}\ .
\ee
The metric must take the form
\be
 ds^2= H^{-1} \left( {\cal {}F} du^2 + 2 du dv \right)-H^2
 \gamma_{mn} dy^m dy^n \ ,
\ee
where the quantities $H$, ${\cal F}$, and $\gamma_{mn}$
depend on $u$ and $y^m$ only (because $V$ is Killing).
It is particularly useful to introduce a null basis
\be
\label{milo}
 e^+ = V = H^{-1} du, \qquad e^- = dv + \frac{1}{2} {\cal {}F} du, \qquad
e^i= H {\hat{e}}^i
\ee
satisfying
\be\label{nflip}
ds^2=2e^+e^- -e^ie^i
\ee
where $ {\hat{e}}^i=  {\hat{e}}^i_m dy^m$ is an orthonormal basis
for the 3-manifold with $u$-dependent metric $\gamma_{mn}$;
$\delta_{ij}  {\hat{e}}^i  {\hat{e}}^j =  \gamma_{mn} dy^m dy^n$.

Equations \p{eqn:VdotX} and \p{eqn:VstarX} imply that $X^{(i)}$
can be written
\be
 X^{(i)} = e^+ \wedge L^{(i)}
\ee
where $L^{(i)} = L^{(i)}{}_m e^m$ satisfy $L^{(i)}{}_{m} L^{(j)}{}_{n}
\delta^{mn} = \delta^{ij}$. In fact, by making a
change of basis we can set $L^{(i)}=e^i$, so
\be\label{sgnx}
 X^{(i)} = e^+ \wedge e^i = du \wedge {\hat{e}}^i \ .
\ee
We set $\epsilon_{+-123}= \eta$; $\eta^2=1$.
Then ({\ref{eqn:cclos}})  implies
\bea
\label{eqn:diferent}
du \wedge d {\hat{e}}^1 &=& 0
\nn
du \wedge \big[ d {\hat{e}}^2 - \c (A \wedge {\hat{e}}^3+ \eta{\sqrt{3}
\over 2} H {\hat{e}}^1 \wedge {\hat{e}}^2) \big] &=& 0
\nn
du \wedge \big[ d {\hat{e}}^3 + \c (A \wedge {\hat{e}}^2- \eta{\sqrt{3}
\over 2} H {\hat{e}}^1 \wedge {\hat{e}}^3) \big] &=& 0 \ .
\eea
Now define ${\tilde{d}} {\hat{e}}^i = {1 \over 2} ({\partial {\hat{e}}^i_m
\over \partial y^n} - {\partial {\hat{e}}^i_n \over
\partial y^m}) dy^n \wedge dy^m$. Then ({\ref{eqn:diferent}}) implies that
\bea
\label{eqn:diferentb}
{\tilde{d}} {\hat{e}}^1&=& 0
\nn
{\tilde{d}} {\hat{e}}^2 - \c (A \wedge {\hat{e}}^3+ \eta{\sqrt{3} \over 2}
H {\hat{e}}^1 \wedge {\hat{e}}^2) &=& 0
\nn
{\tilde{d}}  {\hat{e}}^3 + \c (A \wedge {\hat{e}}^2- \eta {\sqrt{3} \over 2}
H {\hat{e}}^1 \wedge {\hat{e}}^3) &=& 0 \ .
\eea
Hence, in particular $({\hat{e}}^2+i {\hat{e}}^3) \wedge {\tilde{d}}
  ({\hat{e}}^2+i {\hat{e}}^3)=0$
from which it follows that there exists a complex function $S(u,y)$
and real functions
$x^2=x^2(u,y)$, $x^3=x^3(u,y)$ such that
\be
({\hat{e}}^2+i {\hat{e}}^3)_m = S {\partial \over \partial y^m} (x^2+ix^3)
\ee
and hence $({\hat{e}}^2+i {\hat{e}}^3) = S d(x^2+ix^3)+ \psi du$ for
some complex function $\psi(u,y)$. Similarly, there exists a real function
 $x^1=x^1(u,y)$ such that ${\hat{e}}^1 = dx^1 + a^1 du$ for some
real function $a^1$. Hence, from this it is clear that we can change
 coordinates from $u,y^m$ to $u,x^m$.
Moreover, we can make a gauge transformation of the form
$A \rightarrow A + d \Lambda$ where $\Lambda=\Lambda(u,x)$ in order to set
$X^2+i X^3 \rightarrow S du \wedge (dx^2+idx^3)$ where $S$ is now a
real function.
Note that such a gauge transformation preserves the original gauge
restriction \p{gaugechoice} that $A_v=0$.

Hence, the null basis can be simplified to
\bea
\label{eqn:milob}
 e^+ = V &=& H^{-1} du, \quad e^- = dv + \frac{1}{2} {\cal {}F} du
 \nn
e^1 = H(dx^1+ a^1 du), \quad e^2 &=& H (Sdx^2 + S^{-1}a^2 du), \quad e^3
= H  (S dx^3 + S^{-1} a^3 du)
\eea
for real functions $H(u,x^m)$, $S(u,x^m)$, $a^i (u,x^m)$,
and $X^i = e^+ \wedge e^i$.

Equation \p{eqn:df} implies that $i_V F=0$ and hence
\be
 {}F = {}F_{+i} e^+ \wedge e^i + \frac{1}{2} {}F_{ij} e^i \wedge e^j \ .
\ee
To proceed, we use ({\ref{eqn:dV}}) to solve for the components $F_{ij}$;
we find
\be
\label{eqn:ffixa}
F_{12} = - \eta {\sqrt{3} \over 2} H^{-2} S^{-1} \nabla_3 H
\ , \quad F_{13} = \eta {\sqrt{3} \over 2} H^{-2} S^{-1} \nabla_2 H
\ , \quad F_{23} = \eta ({\c \over 2} - {\sqrt{3} \over 2} H^{-2} \nabla_1 H)
\ee
where $\nabla$ denotes the flat connection on $\bR^3$, $\nabla_i \equiv
{\partial \over \partial x^i}$, and we set $a^1=a_1$, $a^2=a_2$ and
$a^3=a_3$. Next we consider the
constraints implied by ({\ref{eqn:dPhi}}).  After a long calculation
we find {\footnote{The origin of this fixed orientation is that
we chose a frame such
$X^{(i)} = e^+ \wedge e^i$, as in \p{sgnx}, rather than
$X^{(i)} = -e^+ \wedge e^i$.}
$\eta=-1$ together with
\be
\label{eqn:fixh}
\nabla_1 S = -{\c \sqrt{3} \over 2} HS
\ee
and we also find that the gauge field strength is
\bea
\label{eqn:fldstr}
F &=& \big(-{\c \over \sqrt{3}} HA_u +{1 \over 2 \sqrt{3}}
S^{-2} H^{-2}\big[\nabla_2 (H^3 a_3)- \nabla_3 (H^3 a_2)\big] \big)
du \wedge dx^1
\nn
&-&{1 \over 2 \sqrt{3}} H^{-2} \big[\nabla_1 (H^3 a_3)- \nabla_3
(H^3 a_1)\big]du
\wedge dx^2 + {1 \over 2 \sqrt{3}} H^{-2} \big[\nabla_1 (H^3 a_2)- \nabla_2
(H^3 a_1)\big] du \wedge dx^3
\nn
&+&{\sqrt{3} \over 2} ( \nabla_3 H dx^1 \wedge dx^2 -
\nabla_2 H dx^1 \wedge dx^3)+{1 \over 2}(\sqrt{3} \nabla_1 H - \c H^2)
 S^2 dx^2 \wedge dx^3
\eea
and the gauge field potential is
\be
\label{eqn:gpot}
A = A_u du + {1 \over \c S}(\nabla_2 S dx^3 - \nabla_3 S dx^2) \ .
\ee
We require that $F=dA$, which implies that
\bea
\label{eqn:aueqs}
{1 \over 2 \sqrt{3}} \big[\nabla_2 (H^3 a_3)- \nabla_3 (H^3 a_2) \big] &=&
-H^2 S^{4 \over 3} \nabla_1 (S^{2 \over 3} A_u)
\nn
{1 \over 2 \sqrt{3}} \big[\nabla_3 (H^3 a_1)- \nabla_1 (H^3 a_3) \big] &=&
-H^2 \nabla_2 A_u -{H^2 \over \c} \nabla_3
\big( S^{-1} {\partial S \over \partial u} \big)
\nn
{1 \over 2 \sqrt{3}} \big[\nabla_1 (H^3 a_2)- \nabla_2 (H^3 a_1) \big] &=&
-H^2 \nabla_3 A_u +{H^2 \over \c} \nabla_2
\big( S^{-1} {\partial S \over \partial u} \big)
\eea
and
\be
\label{eqn:seqn}
S \nabla_1 \nabla_1 S -{1 \over 3} (\nabla_1 S)^2 +S^{-1} (\nabla_2 \nabla_2
S+ \nabla_3 \nabla_3 S)-S^{-2}\big[ (\nabla_2 S)^2+ (\nabla_3 S)^2 \big]=0
\ee
where we have made use of ({\ref{eqn:fixh}}) in order to simplify these
equations. Observe that ({\ref{eqn:aueqs}}) implies the following
integrability condition:
\be
\label{eqn:integriii}
\nabla_1 \big[ H^2 S^{4 \over 3} \nabla_1 (S^{2 \over 3} A_u)\big]+
 \nabla_2 (H^2 \nabla_2 A_u)
+ \nabla_3 (H^2 \nabla_3 A_u) ={2H \over \c} \big[ \nabla_3 H
\nabla_2 (S^{-1} {\partial S \over \partial u}) -
\nabla_2 H
\nabla_3 (S^{-1} {\partial S \over \partial u}) \big] \ .
\ee
In fact, it is straightforward to show that these constraints
ensure that the Bianchi identity and the gauge field equations hold
automatically. In addition, all but the $uu$ component of the
Einstein equations also hold automatically. The $uu$ component fixes
${\cal{F}}$ in terms of the other fields.

{}Finally, it remains to substitute the bosonic constraints into the
killing spinor equation ({\ref{eqn:kspin}}) and to check that the geometry
does indeed admit Killing spinors.
If we impose the constraint
\be
\label{eqn:annihksp}
\gamma^+ \epsilon^a =0
\ee
on the Killing spinor,
then the $\alpha=-$ component of the Killing spinor equation implies
that
\be
\label{eqn:liedera}
{\partial \epsilon^a \over \partial v}=0 \ ,
\ee
so $\epsilon^a = \epsilon^a (u,x^1,x^2,x^3)$.
Next we set $\alpha=+$; we find that
\bea
\label{eqn:dplus}
H \big( {\partial \epsilon^a \over \partial u}-
a^1 \nabla_1 \epsilon^a -S^{-2} a^2 \nabla_2 \epsilon^a
-S^{-2} a^3 \nabla_3 \epsilon^a \big)\nn
-{\c \over 4 \sqrt{3}} \gamma^-
 (\gamma^1 \epsilon^a + \epsilon^{ab} \epsilon^b)
+{\chi A_+\over 2}(\gamma^1 \epsilon^a + \epsilon^{ab} \epsilon^b)
 =0 \ .
\eea
Acting on ({\ref{eqn:dplus}}) with $\gamma^+$ we find the algebraic
constraint
\be
\label{eqn:extraconstr}
\gamma^1 \epsilon^a + \epsilon^{ab} \epsilon^b=0 \ .
\ee
Next set $\alpha=1,2,3$; it is straightforward to show that these components
of the Killing spinor equation imply that
\be
\label{eqn:constabc}
\nabla_1 \epsilon^a = \nabla_2 \epsilon^a = \nabla_3 \epsilon^a =0
\ee
and substituting this back into ({\ref{eqn:dplus}}) we also find
\be
\label{eqn:dpplus}
{\partial \epsilon^a \over \partial u} =0 \ .
\ee
Hence the Killing spinor equation implies that $\epsilon^a$ is constant.

It is also useful to examine the effect on the solution of certain
co-ordinate transformations. In particular, under the shift
$v=v' + g(u,x)$ we note that the form of the solution remains the same,
with $v$ replaced by $v'$, and $a_i$ and $\cF$
replaced by
\bea
\label{eqn:coordsimp}
a'{}_i &=& a_i - H^{-3} \nabla_i g
\nn
\cF' &=& \cF +2 {\partial g \over \partial u}-2  (a_1 \nabla_1 g
 +S^{-2}\big[ a_2 \nabla_2 g + a_3 \nabla_3 g\big])
\nn
&+&H^{-3} \big((\nabla_1 g)^2+ S^{-2}\big[(\nabla_2 g)^2+
 (\nabla_3 g)^2\big] \big)
\eea
hence we see that $H^3 a$ is determined only up to a gradient.

To summarize, it is possible to construct a null supersymmetric solution
as follows. First choose $S(u,x)$ satisfying
({\ref{eqn:seqn}}). Then use ({\ref{eqn:fixh}}) to obtain $H$.
Next find $A_u(u,x)$ satisfying ({\ref{eqn:integriii}}).
Given such an $A_u$ the equations ({\ref{eqn:aueqs}}) can always be solved,
at least locally, to give $H^3 a$ up to a gradient;
this gradient term can be removed by making a shift in $v$ as
described above. Then the gauge potential
is given by ({\ref{eqn:gpot}}). Lastly, fix $\cF$ by solving the
$uu$ component of the Einstein equations. In this sense the
solutions are determined by three functions $S,A_u$ and $\cal F$.
The Killing spinors are constant
and constrained by  ({\ref{eqn:annihksp}}) and ({\ref{eqn:extraconstr}}).
Note that these solutions are therefore generically 1/4-supersymmetric,
in contrast with the null solutions in the ungauged supergravity, which
are generically 1/2-supersymmetric.

\subsection{Magnetic String solutions.}

To construct a solution to these equations,
we take $S$ to be independent of $u$ and
separable, $S=P(x^1) Q(x^2,x^3)$, so that from
({\ref{eqn:seqn}}) we find that
\be
\label{eqn:simplera}
\big( \nabla_2 \nabla_2 + \nabla_3 \nabla_3 \big) \log Q = -k Q^2
\ee
and
\be
\label{eqn:simplerb}
P \ddot{P} - {1 \over 3} (\dot{P})^2 -k=0
\ee
for constant $k$, where here $\dot{}={d \over d x^1}$. We then
have $H=-{2 \over \c \sqrt{3}}P^{-1}{\dot{P}}$. We set
$A_u=a^1=a^2=a^3=0$ and seek solutions that also have ${\cal{F}}=0$.
The metric and the
gauge field strength are given by
\be
\label{eqn:mmeta}
ds^2 = - \c \sqrt{3} P (\dot{P})^{-1} du dv -{4 \over 3 \c^2} P^{-2}
(\dot{P})^2 (dx^1)^2 -{4 \over 3 \c^2}(\dot{P})^2 ds^2 (M_2)
\ee
and
\be
\label{eqn:fstra}
F = -k \c^{-1} dvol (M_2)
\ee
where $M_2$ is a 2-manifold with metric
\be
\label{eqn:twomanifld}
ds^2 (M_2) = Q^2 \big[ (dx^2)^2 + (dx^3)^2 \big] \ .
\ee
Because $Q$ satisfies \p{eqn:simplera}, we see that $M_2$ has constant
curvature and hence can be taken to be $\bR^2$ if $k=0$,
$S^2$ if $k>0$ (with radius $k^{-{1 \over 2}}$), or
$H^2$ if $k<0$ (with radius $(-k)^{-{1 \over 2}}$).
Next we simplify the metric by defining $R={\dot{P}}$, and we note that
({\ref{eqn:simplerb}}) implies that $R= \sqrt{\mu P^{2 \over 3} -3k}$ for
constant $\mu$ and also $R^2 ({R^2 \over 3}+k)^{-2} dR^2 =
 P^{-2} {\dot{P}}^2 (dx^1)^2$.
Hence, on re-scaling ${\hat{v}}= -9 \c \mu^{-{3 \over 2}} v$ we obtain
\be
\label{eqn:evensimplermet}
ds^2 = R^{1 \over 2} ({R \over 3}+{k \over R})^{3 \over 2} du d {\hat{v}}
-{4 \over 3 \c^2} ({R \over 3}+{k \over R})^{-2} dR^2
-{4 \over 3 \c^2}R^2 ds^2 (M_2) \ .
\ee
It is straightforward to show that all components of the Einstein
equations are satisfied. These solutions are the  black string
solutions of \cite{chamsab,sabrahyp}. {}For $k<0$ the solution has a
horizon
at $R^2=-3k$ and the near horizon limit gives $AdS_3\times H^2$, which
we also found in the timelike class of solutions.

\sect{Integrability and Maximal Supersymmetry}

The Killing spinor equation ({\ref{eqn:kspin}})
implies the following integrability conditions on the Killing spinor:
\bea
\label{eqn:integrab}
\frac{1}{8}{}^5 R_{\rho\mu\nu_1\nu_2}\gamma^{\nu_1\nu_2}\epsilon^a
&=&-\frac{1}{4{\sqrt 3}}(\gamma{_{[\mu}}{^{\nu_1\nu_2}}
+4\gamma^{\nu_1}\delta_{[\mu}^{\nu_2})
\nabla_{\rho]}{}F_{\nu_1\nu_2}\epsilon^a \nn
&+&\frac{1}{48}(-2{}F^2\gamma_{\mu\rho}
-8{}F^2_{\nu [\rho}\gamma^\nu{}_{\mu]}
+12{}F_{\mu\nu_1}{}F_{\rho\nu_2}\gamma^{\nu_1\nu_2}
+8{}F_{\nu_1\nu_2}{}F_{\nu_3 [\rho}\gamma_{\mu]}{}^{\nu_1\nu_2\nu_3}
) \epsilon^a
\nn
&+&{\c \over 24} \big(\gamma_{\rho \mu}{}^{\nu_1 \nu_2} F_{\nu_1
\nu_2}
-4 F_{\nu [\rho} \gamma_{\mu ]}{}^{\nu}
-6 F_{\rho \mu} \big) \epsilon^{ab} \epsilon^b +{\c^2 \over 48}
 \gamma_{\rho \mu} \epsilon^a \ .
\eea

To obtain a geometry preserving maximal supersymmetry, we
require that this integrability condition imposes no
algebraic constraints on the Killing spinor.
In particular, it is required that the terms which are zeroth, first and
second order in the gamma-matrices should vanish independently
(after rewriting the terms cubic, quartic and quintic in
gamma-matrices in terms of quadratic, linear and
zeroth order terms, respectively).
Hence from the zeroth order term we immediately obtain $F=0$.
The integrability condition then simplifies considerably to give
\be
\label{eqn:simplerint}
{}^5 R_{\rho\mu\nu_1\nu_2}={\c^2 \over 12} \big( g_{\rho \nu_1}
g_{\mu \nu_2}-g_{\rho \nu_2} g_{\mu \nu_1} \big) \ ,
\ee
which implies that that the five dimensional geometry must be $AdS_5$.
This is in contrast to the case of the ungauged theory, for which it
has been shown \cite{gaunguthull}
that there is a rich structure of maximally supersymmetric
solutions.

Note also that if we contract the integrability
condition with $\gamma^\mu$ we get
\bea
0&=&\left(R_{\rho\mu}+2F_{\rho\nu}F_{\mu}{}^\nu-\frac{1}{3}g_{\rho\mu}
(F^2+\chi^2) \right)\gamma^{\mu}\e\nn
&-&\frac{1}{\sqrt 3}\left[
*(d*{}F+\frac{2}{\sqrt 3}{}F \wedge
{}F)\right] ^{\nu}(2g_{\nu\rho}-\gamma_{\rho\nu})\e\nn &-&\frac{1}
{6{\sqrt 3}}d{}F_{\nu_1\nu_2\nu_3}(\gamma_{\rho}{^{\nu_1\nu_2\nu_3}}
-6\delta_{\rho}^{\nu_1}\gamma^{\nu_2\nu_3})\e \ .
\eea
Suppose we have a geometry admitting a Killing spinor
and in addition the equation of motion and Bianchi identity for $F$
are satisfied. By following exactly the same argument presented in
\cite{gaunguthull} we conclude that if the Killing spinor is
timelike, then all of Einstein's equations are automatically satisfied
while if it is null, the $++$ component, in the frame \p{nflip},
might not be satisfied.

\sect{Conclusions}

In this paper we have presented a classification of all supersymmetric
solutions of minimal five-dimensional gauged supergravity.
One of the interesting differences with the ungauged theory
is that in the timelike case much more of the solution is fixed by
the geometric structure of the base manifold. On the other hand,
in the gauged case the base must be K\"ahler and not hyper-K\"ahler,
whereas in the ungauged case the base must be hyper-K\"ahler.
In the null case the solutions are still determined by three differential
equations as in the ungauged case, but these equations are more
complicated than those in the ungauged theory. In addition we have shown
that the gauging generically reduces the proportion of supersymmetry
preserved from 1/2 to 1/4.  In the gauged theory, $AdS_5$
is the unique maximally supersymmetric solution, while there are
a number of different possibilities in the ungauged case.

We have also presented some new
solutions, that would be worth investigating further both in $d=5$
and in $d=10$ after uplifting with a five-sphere.
Many of the new solutions we have presented have closed timelike curves, as
was seen in the ungauged case, which provides additional evidence that
they are a commonplace amongst supersymmetric solutions. It would
be interesting to see if they can be removed in our solutions by going
to a covering space either in five or ten dimensions.
Moreover, all of the timelike solutions which we have examined correspond
to K\"ahler geometries for which the tensor
$T$ given by ({\ref{eqn:tdef}}) vanishes. Clearly, there are many new
solutions
for which $T \neq 0$.

It may also be possible to use the generic form of the supersymmetric
solutions to examine the geometry of black hole solutions. In
\cite{Reall:2002bh}, the constraints on ungauged solutions found in
\cite{gaunguthull} were used to show that the near horizon geometry of
all supersymmetric black holes is isometric to the near horizon
geometry of the BMPV solutions; and from this a uniqueness theorem was
proven. In contrast, it is known that the static asymptotically
anti-de-Sitter black holes have no horizon, as they are nakedly singular.
However, there does exist a class of rotating AdS black hole solutions
which have horizons, and hence a similar investigation could be feasible.

\subsection*{Acknowledgments}
J.B.G. thanks EPSRC for support. We would like to thank Dario Martelli
and Harvey Reall for useful discussions.


\begin{thebibliography}{99}

\bibitem{gaunguthull}
J. P. Gauntlett, J. B. Gutowski, C. M. Hull, S. Pakis and H. S. Reall,
{\it{All Supersymmetric Solutions of Minimal Supergravity in
Five-Dimensions}}; hep-th/0209114.

\bibitem{Gauntlett:2002fz}
J.~P.~Gauntlett and S.~Pakis,
{\it{The geometry of D = 11 Killing spinors}};
hep-th/0212008.

\bibitem{tod}
K.~P. Tod, {\it{All Metrics Admitting Supercovariantly Constant Spinors}},
Phys. Lett. {\bf{B121}} (1983) 241; {\it{More on Supercovariantly
Constant Spinors}}, Class. Quant. Grav. {\bf{12}} (1995) 1801.

\bibitem{Friedrich:2001nh}
T.~Friedrich and S.~Ivanov,
{\it{Parallel spinors and connections with skew-symmetric torsion in string
theory}}, Asian Journal of Mathematics {\bf{6}} (2002) 303; math.DG/0102142;
{\it{Killing spinor equations in dimension 7 and geometry of integrable
$G_2$-manifolds}}; math.DG/0112201.

\bibitem{Gauntlett:2002sc}
J.~P.~Gauntlett, D.~Martelli, S.~Pakis and D.~Waldram,
{\it G-structures and wrapped NS5-branes}; hep-th/0205050.

\bibitem{Chamblin:1999tk}
A.~Chamblin, R.~Emparan, C.~V.~Johnson and R.~C.~Myers,
{\it{Charged AdS black holes and catastrophic holography}},
Phys.\ Rev.\ D {\bf 60} (1999) 064018; hep-th/9902170.

\bibitem{Cvetic:1999xp}
M.~Cvetic {\it et al.},
{\it{Embedding AdS black holes in ten and eleven dimensions}},
Nucl.\ Phys.\ B {\bf 558} (1999) 96; hep-th/9903214.

\bibitem{Gunaydin:1983bi}
M.~Gunaydin, G.~Sierra and P.~K.~Townsend,
{\it{The Geometry Of N=2 Maxwell-Einstein Supergravity And Jordan
Algebras}}, Nucl.\ Phys.\ B {\bf 242} (1984) 244.

\bibitem{Reall:2002bh}
H.~S.~Reall,
{\it{Higher dimensional black holes and supersymmetry}}
;hep-th/0211290.

\bibitem{popeqg}
C. N. Pope, {\it{K\"ahler manifolds and quantum gravity}},
J. Phys. {\bf{A15}} (1982) 2455.

\bibitem{rbpstrom}
R. Britto-Pacumio, A. Strominger and A. Volovich, {\it{Holography for Coset
Spaces}}, JHEP 9911 (1999) 013; hep-th/9905211.

\bibitem{sabrahyp}
D. Klemm and W. A. Sabra, {\it{Supersymmetry of Black Strings in D=5
Gauged Supergravities}}, Phys. Rev. {\bf{D62}} (2000) 024003;
hep-th/0001131.

\bibitem{gubconf}
S. Gubser, {\it{Einstein Manifolds and Conformal Field
Theories}}, Phys. Rev. {\bf{D59}} 025006 (1999);
hep-th/9807164.

\bibitem{londi}
L. London, {\it{Arbitrary Dimensional Cosmological Multi-Black Holes}},
Nucl. Phys. {\bf{B434}} (1995) 709.

\bibitem{sabbh}
D. Klemm and W. A. Sabra, {\it{Charged Rotating Black Holes
in 5-D Einstein-Maxwell (A)DS Gravity}}, Phys. Lett. {\bf{B503}}
(2001) 147; hep-th/0010200.

\bibitem{chamsab}
A. H. Chamseddine and W. A. Sabra, {\it{Magnetic Strings in
Five-Dimensional Gauged Supergravity Theories}},
Phys. Lett. {\bf{B477}} (2000) 329; hep-th/9911195.









\end{thebibliography}
\end{document}